\documentclass[runningheads]{llncs}
\usepackage{graphicx}
\usepackage{booktabs}
\usepackage{cite}
\usepackage{amsmath,amssymb,amsfonts}
\usepackage{algorithmic}
\usepackage{textcomp}
\usepackage{xcolor}
\usepackage{booktabs}
\usepackage{multirow}
\usepackage{graphicx}
\usepackage{lipsum,lineno} 
\usepackage{float}

\begin{document}
\title{Teachers' perspective on fostering computational thinking through educational robotics\thanks{This research was supported by the NCCR Robotics, Switzerland. We would like to thank the trainers and teachers who made this study possible.}}
\titlerunning{Teachers' perspective on fostering CT through educational robotics}
\author{
Morgane Chevalier\inst{1,2,}\thanks{Morgane Chevalier and Laila El-Hamamsy contributed equally to this work}\orcidID{0000-0002-9115-1992} \and
Laila El-Hamamsy\inst{1,\star\star}\orcidID{0000-0002-6046-4822} \and
Christian Giang\inst{1,3}\orcidID{0000-0003-2034-9253} \and
Barbara Bruno\inst{1}\orcidID{0000-0003-0953-7173} \and
Francesco Mondada\inst{1}\orcidID{0000-0001-8641-8704}
}
\authorrunning{M. Chevalier \& L. El-Hamamsy et al.}
\institute{Ecole Polytechnique Fédérale de Lausanne (EPFL), Switzerland \\
\email{firstname.lastname@epfl.ch}
\and
Haute École Pédagogique du canton de Vaud (HEP-VD), Lausanne, Switzerland
\and
SUPSI-DFA, Locarno, Switzerland
}

\maketitle
\begin{abstract}
With the introduction of educational robotics (ER) and computational thinking (CT) in classrooms, there is a rising need for operational models that help ensure that CT skills are adequately developed. One such model is the Creative Computational Problem Solving Model (CCPS) which can be employed to improve the design of ER learning activities. Following the first validation with students, the objective of the present study is to validate the model with teachers, specifically considering how they may employ the model in their own practices. The Utility, Usability and Acceptability framework was leveraged for the evaluation through a survey analysis with 334 teachers. 
Teachers found the CCPS model \emph{useful} to foster transversal skills but could not recognise the impact of specific intervention methods on CT-related cognitive processes. Similarly, teachers perceived the model to be \emph{usable} for activity design and intervention, although felt unsure about how to use it to assess student learning and adapt their teaching accordingly. Finally, the teachers \emph{accepted} the model, as shown by their intent to replicate the activity in their classrooms, but were less willing to modify it or create their own activities, suggesting that they need time to appropriate the model and underlying tenets. 

\ifdefined\USEIEEE
\else 
\fi 
\end{abstract}

\section{Introduction}

Educational robotics has garnered significant interest in recent years to teach students not only the fundamentals of robotics, but also core Computer Science (CS) concepts \cite{el-hamamsy_computer_2020} and Computational Thinking (CT) competencies \cite{bers_computational_2014, chevalier_fostering_2020}. 
However, participation in an ER Learning Activity (ERLA) does not automatically ensure student learning \cite{siegfried_improved_2017}, with the design of the activity playing a key role towards the learning outcomes \cite{fanchamps_influence_2019}.
Indeed, the lack of understanding as to how specific instructional approaches impact student learning in ER activities has been raised at multiple occasions \cite{sapounidis_educational_2020}. 
Many researchers have even evoked the need to have an operational model to understand how to foster CT skills \cite{lye_review_2014} within the context of ER activities \cite{atmatzidou_advancing_2016, ioannou_exploring_2018}.  
To that effect, Chevalier \& Giang et al. \cite{chevalier_fostering_2020} developed the Creative Computational Problem Solving (CCPS) model for CT competencies using an iterative design-oriented approach \cite{rogers2011interaction} through student observations. The resulting 5 phase model (Fig. \ref{fig:lawnmower_activity}) helped identify and understand students’ cognitive processes while engaging in ER learning activities aimed to foster CT skills. 
By analysing the students' behaviour through the lens of the CCPS model to understand the students’ thought processes, both teachers and researchers may have a means of action and intervention in the classroom to foster the full range of cognitive processes involved in creative computational problem solving. 
The authors concluded that a validation by teachers was essential to ensure that they could ``effectively take advantage of the model for their teaching activities'', not only at the design stage, but also to guide specific interventions during ER learning activities.

This article reports the findings of a study involving 334 in-service and pre-service primary school teachers, with the purpose of evaluating their perception of the model and investigating whether their own needs as users of the model are met. 
The Utility, Usability and Acceptability framework \cite{tricot_utility_2003} for computer-based learning environment assessment was leveraged, as it has been previously used for the evaluation of teachers’ perception of the use of educational robots in formal education \cite{chevalier_pedagogical_2016}. 
More formally, we address the following questions:

\textbf{RQ1:} What is the perceived utility of the CCPS model?

\textbf{RQ2:} What is the perceived usability of the model?

\textbf{RQ3:} What is the acceptability of the model by teachers?

\section{Methodology} 
\label{sec:methodology}

\begin{figure}
    \centering
    \vspace{-5pt}
    \includegraphics[height=0.18\textheight]{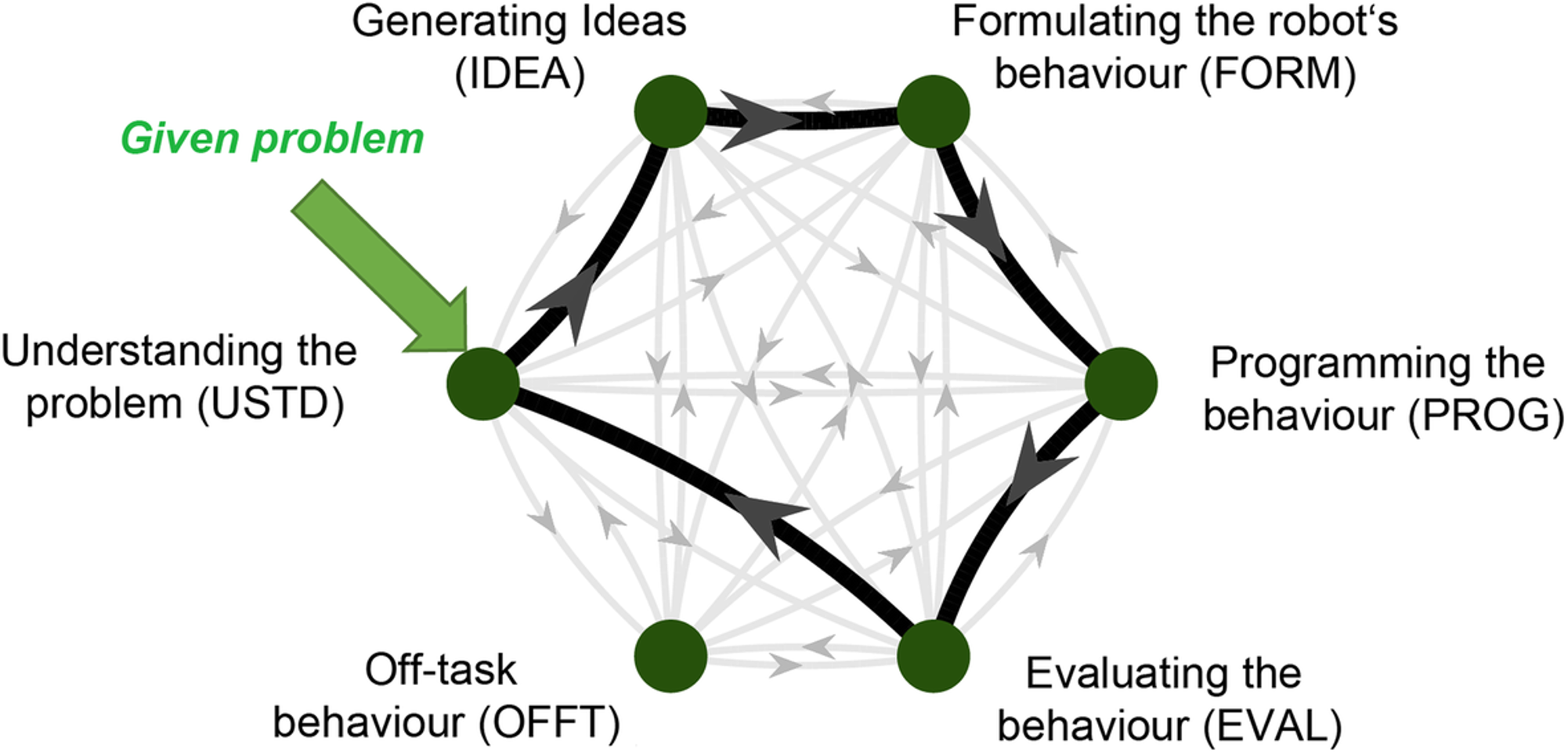}
    \includegraphics[height=0.18\textheight]{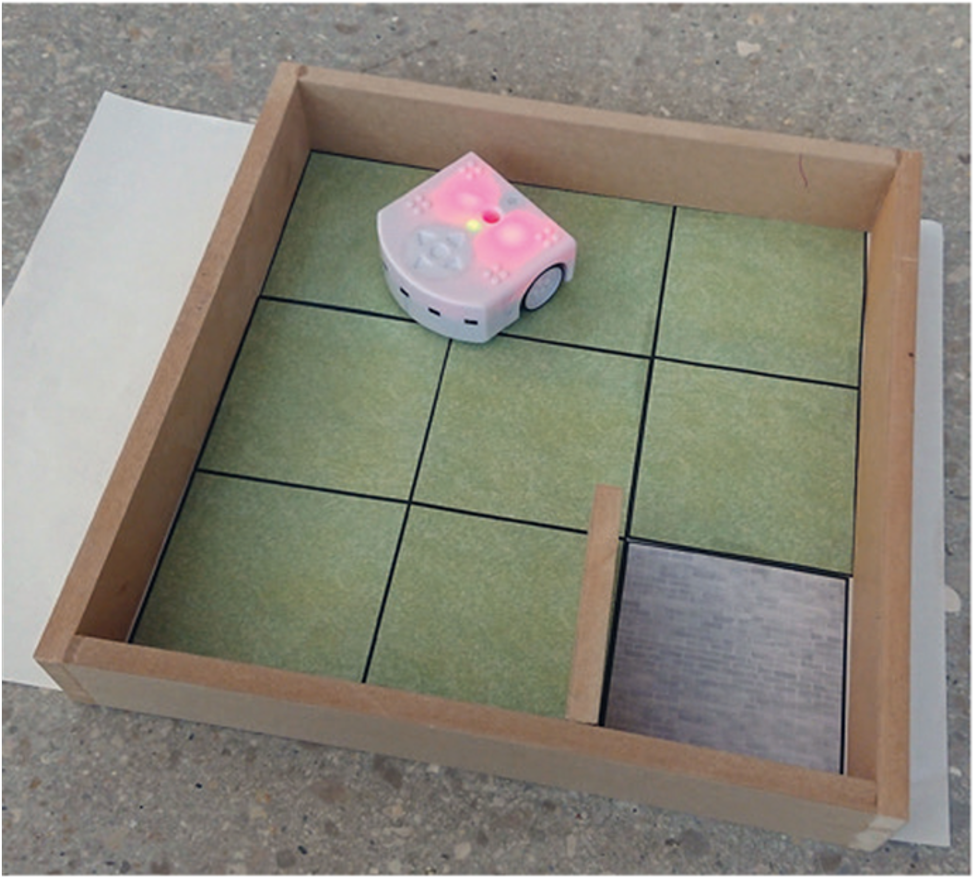}
    \vspace{-5pt}
    \caption{The CCPS model (left) and lawnmower activity setup with Thymio (right) \cite{chevalier_fostering_2020}.
    }
    \vspace{-5pt}
    \label{fig:lawnmower_activity}
\end{figure}

To evaluate the model, the study was conducted with 232 in-service and 102 pre-service teachers participating in the mandatory training program for Digital Education underway in the Canton Vaud, Switzerland \cite{el-hamamsy_computer_2020} between November 2019 and February 2020. The inclusion of both pre-service and in-service teachers within the context of a mandatory ER training session helps ensure the generalisability of the findings to a larger pool of teachers, and not just experienced teachers and/or pioneers who are interested in ER and/or already actively integrating ER into their practices \cite{chevalier_pedagogical_2016}. 
During the ER training session, the teachers participated in an ER learning activity (see Lawnmower activity in Fig. \ref{fig:lawnmower_activity} \cite{chevalier_fostering_2020}) which was mediated by the CCPS model. 
During this activity, the teachers worked in groups of 2 or 3 to program the event-based Thymio II robot \cite{mondada_bringing_2017} to move across all of the squares in the lawn autonomously.
As the robot is event-based, the participants ``have to reflect on how to use the robot’s sensors and actuators to generate a desired [behaviour]'' \cite{chevalier_fostering_2020}, which requires that the participants leverage many CT-related competencies. 
So that teachers understand how the CCPS can be used to mediate an ER learning activity, and similarly to Chevalier \& Giang et al. (2020) \cite{chevalier_fostering_2020}, a temporary access blocking to the programming interface was implemented at regular time intervals.  

After the activity, the teachers participated in a debriefing session where they were asked to express what they had to do to solve the problem and the trainer grouped these comments into categories relating to CT, the CCPS model and transversal skills. 
The teachers were then presented the CCPS model itself and its 5 phases, together with the results of the study conducted by Chevalier \& Giang et al. \cite{chevalier_fostering_2020} to provide concrete testimony as to the effectiveness of the model when applied in classrooms. Finally, in an overarching conclusion about fostering CT competencies during ER activities, the trainer provided guidelines on how to design ER learning activities and intervene accordingly.

The Utility, Usability and Acceptability framework \cite{tricot_utility_2003} then served as the basis for the teachers' evaluation of the CCPS model. As utility ``measures the conformity of the purpose of the device with the users’ needs'' \cite{chevalier_pedagogical_2016, tricot_utility_2003}, in the present context we consider utility with respect to student learning. Two perspectives are adopted : 1) how the use of the model helps foster transversal skills that are part of the mandatory curriculum\footnote{See transversal skills of the curriculum: \url{plandetudes.ch/capacites-transversales}} and 2) how certain intervention methods may help promote reflection in the different phases of the model. Usability on the other hand considers ``the ease of use and applicability of the device'' \cite{chevalier_pedagogical_2016, tricot_utility_2003} by the teacher, which is why the CCPS model in this case is considered in accordance with the ``professional and technical actions'' that teachers make use of in their daily practices\footnote{See ``gestes professionels'': \url{go.epfl.ch/hepvd\_referentiel\_competences\_2015}}. Finally, acceptability ``measures the possibility of accessing the device and deciding to use it, the motivation to do so, and the persistence of use despite difficulties'' \cite{chevalier_pedagogical_2016, tricot_utility_2003}. In this case, we consider acceptability with respect to what the teachers intend to do with the model with increasing levels of appropriation. 
To measure the aforementioned constructs, a set of questions pertaining to each dimension was developed (see Table \ref{tab:survey_items}) with most responses being provided on a 4-point Likert scale (1 - strongly disagree, 2 - disagree, 3 - agree, 4 - strongly agree).

\begin{table}[!ht]
\centering
\vspace{-5pt}
\caption{Utility, usability and acceptability survey \cite{tricot_utility_2003}. Utility in terms of transversal skills considers 5 dimensions: collaboration (\texttt{COL}), communication (\texttt{COM}), learning strategies (\texttt{STRAT}), creative thinking (\texttt{CREA}) and reflexive processes (\texttt{REFL})}
\vspace{-5pt}
\label{tab:survey_items}
\begin{tabular}{p{2cm}p{10cm}}
\toprule
Construct & Question \\ \midrule
\multirow{11}{2cm}{Utility of the CCPS for transversal skills (4-point Likert scale)} & (\texttt{COL}) We exchanged our points of view / evaluated the pertinence of our actions / confronted our ways of doing things  \\
 & (\texttt{COM}) We expressed ourselves in different ways (gestural, written etc…) / identified links between our achievements and discoveries / answered our questions based on collected information \\
 & (\texttt{STRAT}) We persevered and developed a taste for effort / identified success factors / chose the adequate solution from the various approaches \\
 & (\texttt{CREA}) We expressed our ideas in different and new ways / expressed our emotions / were engaged in new ideas and exploited them \\
 & (\texttt{REFL}) We identified facts and verified them / made place for doubt and ambiguity / compared our opinions with each other \\ \midrule
\multirow{5}{2cm}{Utility of the intervention methods (checkboxes)} & What helps i) identify the problem (\texttt{USTD})? ii) generate ideas (\texttt{IDEA})? iii) formulate the solution (\texttt{FORM})? iv) program (\texttt{PROG})? v) evaluate the solution found (\texttt{EVAL})?  \\ 
 & \hspace{10pt}\textit{Max 3 of 5 options :  1) manipulating the robot; 2) writing down the observations; 3) observing 3 times before reprogramming; 4) programming; 5) not being able to program} \\\midrule
\multirow{3}{2cm}{Usability (4-point Likert scale)} & The model helps i) plan an ERLA; ii) intervene during an ERLA; iii) regulate student learning during an ERLA; iv) evaluate student learning during an ERLA \\ \midrule
\multirow{4}{2cm}{Acceptability (4-point Likert scale)} & I will redo the same ERLA in my classroom  \\
 & I will do a similar ERLA that I already know in my classroom \\
 & I will do a similar ERLA that I will create in my classroom \\
 & I will do a more complex ERLA in my classroom \\ \bottomrule
\end{tabular}%
\vspace{-10pt}
\end{table}

\section{Results and Discussion}
\label{sec:results}

\subsection{RQ1 - Utility}

\begin{figure}[ht]
    \centering
    \vspace{-10pt}
    \includegraphics[width=\textwidth]{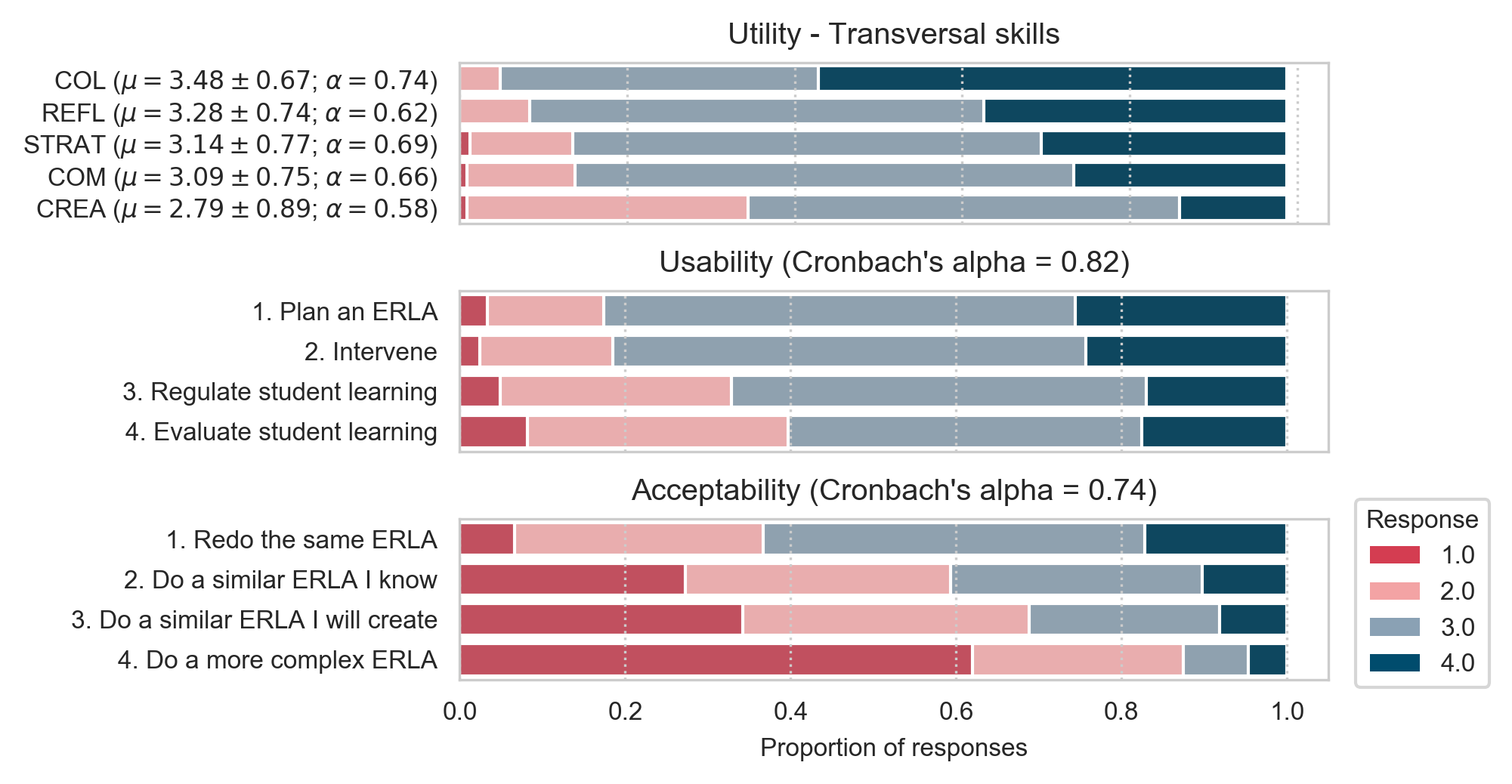}
    \vspace{-20pt}
    \caption{Teachers' perceived utility (with respect to fostering transversal skills), usability and acceptability of the CCPS model. 
    For each transversal skills we report the mean and standard deviation $\mu \pm$std, and Cronbach's $\alpha$ measure of internal consistency.}
    \label{fig:UUA}
    \vspace{-10pt}
\end{figure}

Educational robotics learning activities are often considered to contribute to the development of a number of transversal skills (e.g., collaboration, problem solving etc…) \cite{bers_computational_2014}. While this perception is also shared by teachers who are pioneers in robotics \cite{chevalier_pedagogical_2016}, it is important to ensure that ER activities mediated using the CCPS model and designed to foster CT skills, are perceived by teachers \textit{at large} as contributing to the development of transversal skills. The results of the survey showed that teachers found the ER learning activity with the Thymio useful to engage in transversal skills (Fig. \ref{fig:UUA}), in particular collaboration, reflexive processes, learning strategies and communication, with only creative thinking being less perceived by teachers.
This is coherent with the fact that the ER Lawnmower activity was conceived to promote students' use of transversal skills to help the emergence of related CT competencies and suggests that the use of the CCPS model in designing ER learning activities helps teachers see and strengthen the link between ER, transversal skills and CT, confirming the results of \cite{chevalier_pedagogical_2016} with teachers who were novices. 
Although in the mandatory curriculum teachers are taught to evaluate transversal skills, little indication is provided as to how to foster them. The use of ER learning activities informed and mediated by the CCPS model can provide a concrete way to foster skills already present in the curriculum and ensure that students acquire the desired competencies.

\begin{figure}[ht]
    \centering
    \vspace{-10pt}
    \includegraphics[width=0.8\textwidth]{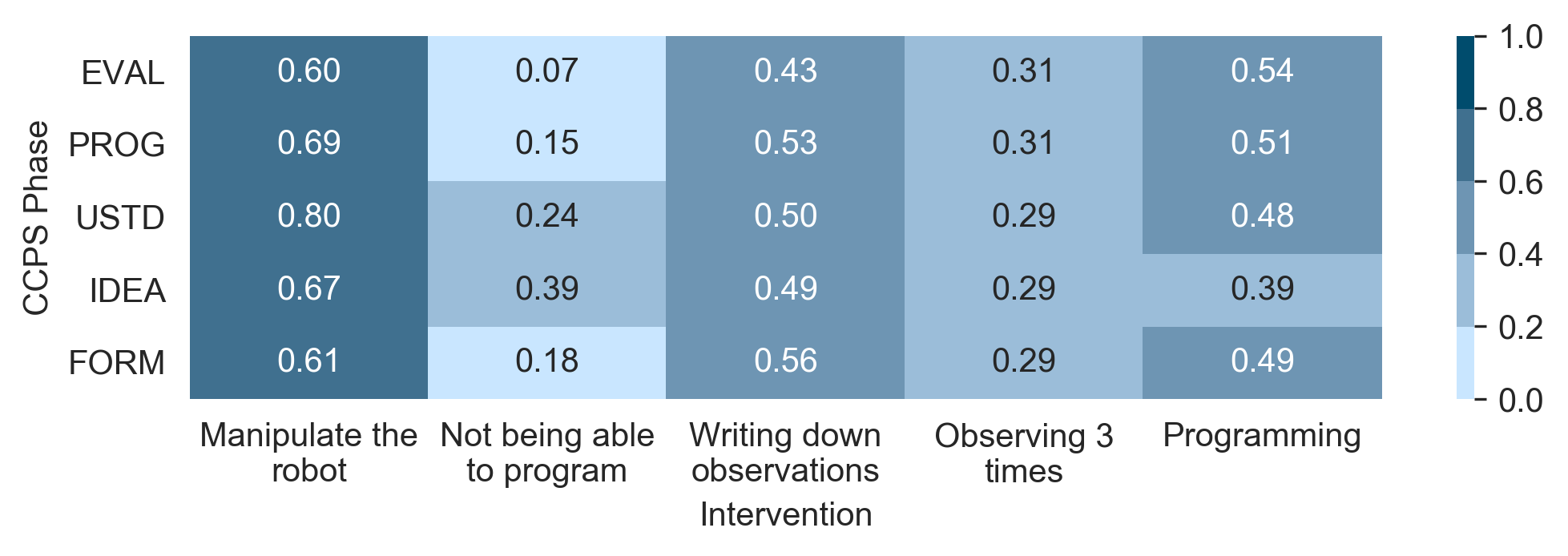}
    \vspace{-15pt}
    \caption{Teachers' perception of the link between intervention methods and the different phases of the CCPS model. For each phase of the model and intervention method, the proportion of teachers having selected the approach as relevant is shown.}
    \vspace{-5pt}
    \label{fig:utility_ccps_intervention}
\end{figure}

To clarify the link between the CCPS model and the employed intervention methods, the teachers were asked to select a maximum of three intervention methods that they believed were useful to engage in each of the phases of the CCPS model (see Fig. \ref{fig:utility_ccps_intervention}). The element which emerges as the most relevant for all the phases of the model is the possibility of manipulating the robot, thereby reinforcing the role of physical agents in fostering CT skills \cite{grover_computational_2013}.
This however is dependent on the fact that the Thymio robot provides immediate visual feedback through the LEDs in relation to each sensor’s level of activation.
This highlights once more the importance of constructive alignment in ER learning activity design \cite{giang_towards_2020} which stipulates the importance of the robot selection in relation to the desired learning outcomes. The second most popular choice was to write down the observations, likely because this constitutes a means of specifying what happens with the robot in the environment. The written observations then become a ``thinking tool'' that supports modelling and investigation \cite{sanchez2008quelles}. 
Surprisingly, and although the teachers were introduced to the fact that unregulated access to the programming interface tends to lead to trial and error behaviour \cite{chevalier_fostering_2020}, programming was often selected as being useful to foster the different CT phases, whilst not being able to program was one of the least selected. Only in the case of idea generation did both programming and not programming receive an equal number of votes. We believe that the frequent selection of programming is due to the fact that the question was based on their experience as the participants, and therefore the need for a high sense of controllability \cite{rolland_motivation_1994}, in the ER learning activity and not on their experience as a teacher leading the activity in the classroom. 

To summarise, on the one hand, teachers perceive the usefulness of promoting transversal skills that they are familiar with, as they are already part of the curriculum. On the other hand, they do not perceive what research has shown to be useful to promote CT competencies, likely because it was not part of the curriculum until now. Therefore, experimentation in their classrooms is necessary, as well as further training to help them acquire a more critical view of ER learning activities and to understand the impact of specific intervention methods on the development of students' skills.

\subsection{RQ2 - Usability}

With respect to the usability of the CCPS model in the teaching profession, responses were globally positive (see Fig. \ref{fig:UUA}, $\mu = 2.89$, std$= 0.78$, Cronbach’s $\alpha = 0.82$). Teachers believed that the CCPS model could be used to plan and intervene during an ER learning activity (over 80\% of positive responses), likely due to the guidelines provided during the theoretical presentation. However, the link between the CCPS model and student learning was less evident for teachers: 66\% believed it could be used to regulate student learning and 60\% that it could be used to evaluate student learning. Both constructs are related by the need to assess students and understand where they stand in terms of the overall learning objectives. Although this shows that teachers need to be taught how to identify the phases in which the students are to be able to use the CCPS model to its full extent, this is also highly linked to the difficulty found in both the ER and CT literature in terms of assessment of learning and/or transversal skills \cite{grover_computational_2013, ioannou_exploring_2018}. 

\subsection{RQ3 - Acceptability}

The question of acceptability here targets teachers’ intent to use the CCPS model in their practices. Intent to use is considered at progressive levels of appropriation (see Fig. \ref{fig:UUA}), which is why it is not surprising to find that while teachers might be willing to conduct the same ER learning activity in their classrooms (64\%), less are willing to adapt the activity (40\%), create their own custom one (32\%), and conduct a more complex one (20\%). One can put this in relation with the Use-Modify-Create (UMC) progression \cite{lytle_use_2019} which was developed to scaffold student learning in CT contexts. 
Teachers need to start by using the model in a given ER learning activity to gain in self-efficacy. Only then will they be able to progress to the next stage where they feel comfortable adapting the use of the model to their teaching style and to the individual students. Finally, teachers will reach a level where they create their own ER pedagogical interventions to foster CT competencies. One must however note that intent is likely influenced by external factors (e.g. time or access to the robots, frequent barriers to the introduction of ER in formal education \cite{chevalier_pedagogical_2016, el-hamamsy_computer_2020}).

\section{Conclusion}

Provided the prominent role that teachers play in the integration of ER and CT in formal education, this study investigated teachers’ perception of an operational model to foster Computational Thinking (CT) competencies through ER activities: the Creative Computational Problem Solving (CCPS) model \cite{chevalier_fostering_2020}. Three research questions were considered in a study with 334 pre-service and in-service primary school teachers : What is the perceived utility (RQ1), usability (RQ2) and acceptability (RQ3) of the CCPS model? 
While teachers found that the activity design and intervention methods employed were useful to foster transversal skills (RQ1), their perception of the utility of the intervention methods on the different cognitive processes defined by the CCPS model (RQ1) was somewhat unexpected. 
In terms of the usability (RQ2), teachers perceived how they could design an activity and intervene using the model, but were less able to perceive how the model could be used to assess where the students were in terms of learning and regulate the activity to mediate their learning. 
The findings of RQ1 and RQ2 support the importance of training teachers to recognise and understand the different cognitive processes to intervene adequately and be able to differentiate their teaching per student, rather than adopting a unique strategy for an entire class. 

To help teachers implement ER learning activities in the classroom and gain in autonomy to create their own activities that foster CT skills (RQ3), it seems relevant to alternate between experimentation in classrooms and debriefing during teacher training and go beyond providing pedagogical resources. 
To conclude, the operationalisation of ER to foster CT skills must also consider the key role that teachers have to play in the introduction of any such model and its application in formal education. 

\bibliographystyle{splncs04}
\bibliography{biblio.bib}

\begin{thebibliography}{10}
\providecommand{\url}[1]{\texttt{#1}}
\providecommand{\urlprefix}{URL }
\providecommand{\doi}[1]{https://doi.org/#1}

\bibitem{atmatzidou_advancing_2016}
Atmatzidou, S., Demetriadis, S.: Advancing students’ computational thinking
  skills through educational robotics: {A} study on age and gender relevant
  differences. Robotics and Autonomous Systems  \textbf{75},  661--670 (Jan
  2016)

\bibitem{bers_computational_2014}
Bers, M.U., Flannery, L., Kazakoff, E.R., Sullivan, A.: Computational thinking
  and tinkering: {Exploration} of an early childhood robotics curriculum.
  Computers \& Education  \textbf{72},  145--157 (Mar 2014)

\bibitem{chevalier_fostering_2020}
Chevalier, M., Giang, C., Piatti, A., Mondada, F.: Fostering computational
  thinking through educational robotics: a model for creative computational
  problem solving. International Journal of STEM Education  \textbf{7}(1), ~39
  (Dec 2020)

\bibitem{chevalier_pedagogical_2016}
Chevalier, M., Riedo, F., Mondada, F.: Pedagogical {Uses} of {Thymio} {II}:
  {How} {Do} {Teachers} {Perceive} {Educational} {Robots} in {Formal}
  {Education}? IEEE Robotics \& Automation Magazine  \textbf{23}(2),  16--23
  (Jun 2016)

\bibitem{el-hamamsy_computer_2020}
El-Hamamsy, L., Chessel-Lazzarotto, F., Bruno, B., Roy, D., Cahlikova, T.,
  Chevalier, M., Parriaux, G., Pellet, J.P., Lanarès, J., Zufferey, J.D.,
  Mondada, F.: A computer science and robotics integration model for primary
  school: evaluation of a large-scale in-service {K}-4 teacher-training
  program. Educ Inf Technol  (Nov 2020)

\bibitem{fanchamps_influence_2019}
Fanchamps, N.L.J.A., Slangen, L., Hennissen, P., Specht, M.: The influence of
  {SRA} programming on algorithmic thinking and self-efficacy using {Lego}
  robotics in two types of instruction. ITDE  (Dec 2019)

\bibitem{giang_towards_2020}
Giang, C.: Towards the alignment of educational robotics learning systems with
  classroom activities p.~176 (2020)

\bibitem{grover_computational_2013}
Grover, S., Pea, R.: Computational {Thinking} in {K}–12: {A} {Review} of the
  {State} of the {Field}. Educational Researcher  \textbf{42}(1),  38--43 (Jan
  2013)

\bibitem{ioannou_exploring_2018}
Ioannou, A., Makridou, E.: Exploring the potentials of educational robotics in
  the development of computational thinking: {A} summary of current research
  and practical proposal for future work. Educ Inf Technol  \textbf{23}(6),
  2531--2544 (Nov 2018)

\bibitem{lye_review_2014}
Lye, S.Y., Koh, J.H.L.: Review on teaching and learning of computational
  thinking through programming: {What} is next for {K}-12? Computers in Human
  Behavior  \textbf{41},  51--61 (Dec 2014)

\bibitem{lytle_use_2019}
Lytle, N., Cateté, V., Boulden, D., Dong, Y., Houchins, J., Milliken, A.,
  Isvik, A., Bounajim, D., Wiebe, E., Barnes, T.: Use, {Modify}, {Create}:
  {Comparing} {Computational} {Thinking} {Lesson} {Progressions} for {STEM}
  {Classes}. In: ITiCSE. pp. 395--401. ACM (2019)

\bibitem{mondada_bringing_2017}
Mondada, F., Bonani, M., Riedo, F., Briod, M., Pereyre, L., Retornaz, P.,
  Magnenat, S.: Bringing {Robotics} to {Formal} {Education}: {The} {Thymio}
  {Open}-{Source} {Hardware} {Robot}. IEEE Robotics Automation Magazine
  \textbf{24}(1),  77--85 (Mar 2017)

\bibitem{rogers2011interaction}
Rogers, Y., Sharp, H., Preece, J.: Interaction design: beyond human-computer
  interaction. John Wiley \& Sons (2011)

\bibitem{rolland_motivation_1994}
Rolland, V.: La motivation en contexte scolaire / {Rolland} {Viau}. Pédagogies
  en développement {Problématiques} et recherches, De Boeck Université
  (1994)

\bibitem{sanchez2008quelles}
Sanchez, {\'E}.: Quelles relations entre mod{\'e}lisation et investigation
  scientifique dans l’enseignement des sciences de la terre? {\'E}ducation et
  didactique (2-2),  93--118 (2008)

\bibitem{sapounidis_educational_2020}
Sapounidis, T., Alimisis, D.: Educational robotics for {STEM}: {A} review of
  technologies and some educational considerations. p.~435 (Dec 2020)

\bibitem{siegfried_improved_2017}
Siegfried, R., Klinger, S., Gross, M., Sumner, R.W., Mondada, F., Magnenat, S.:
  Improved {Mobile} {Robot} {Programming} {Performance} through {Real}-time
  {Program} {Assessment}. In: ITiCSE. pp. 341--346. ACM (Jun 2017)

\bibitem{tricot_utility_2003}
Tricot, A., Plégat-Soutjis, F., Camps, J., Amiel, A., Lutz, G., Morcillo, A.:
  Utility, usability, acceptability: interpreting the links between three
  dimensions of the evaluation of the computerized environments for human
  training ({CEHT}). Environnements Informatiques pour l’Apprentissage Humain
   \textbf{2003} (2003)

\end{thebibliography}

\end{document}